\documentclass[aps,prd,preprint,showpacs,showkeys]{revtex4}
\usepackage{bm}
\usepackage{amsmath}
\usepackage{amssymb}
\usepackage{tikz}
\begin{document}

\title{Spin Hall effect of light in inhomogeneous axion field}

\author{Mansoureh Hoseini and Mohammad Mehrafarin}
\email{mehrafar@aut.ac.ir}
\affiliation{Physics Department, Amirkabir University of Technology, Tehran
15914, Iran}

\begin{abstract}
We study the spin transport of light in weakly inhomogeneous axion field  in a flat Robertson-Walker universe and derive the  spin Hall effect for circularly polarized rays. Regarding primordial quantum fluctuations of the axion field in the de Sitter phase as the origin of the inhomogeneity, we show that the conformal invariance of the correlator determines the root-mean-square (r.m.s) fluctuations of the path of circularly polarized cosmic rays. We explain how the r.m.s fluctuations can be experimentally determined.
\end{abstract}


\maketitle

\section{Introduction}

Pseudoscalar axions were postulated in 1977 by Peccei and Quinn \cite{Peccei1977} in order to solve the problem of strong CP-invariance. These and other very light axion-like particles, which have become leading candidates for the missing matter of Universe \cite{Kolb1990,Marsh2016}, have been long searched for since their proposal (see, e.g., \cite{Redondo2011,Graham2015,Irastorza2018}).
In the context of cosmology, axions appear after a phase transition in the early Universe at very high energies, when the PQ symmetry breaks spontaneously \cite{Kolb1990}. In the pre-inflation scenario for axions, this symmetry breaking occurs during inflation and the exponential expansion smooths the field, leaving only small spatial perturbations at the end.

The theory of interaction between electromagnetic and pseudoscalar fields was elaborated by Ni in 1977~\cite{Ni1977}. Its well-known application is the prediction of polarization rotation effect (cosmological birefringence) \cite{Carroll1990,Carroll1991,Harari1992} for linearly polarized light propagating over cosmological distances, most notably, for the CMB radiation (see, e.g., \cite{Wu2009,Chiang2010,Jarosik2011}). We have recently shown \cite{Hoseini2019} that this polarization plane rotation arises from an adiabatic geometric phase that appears in the quantum state of photons interacting with the background axion field. 

The present work is motivated by the fact that inhomogeneity in propagation medium generally leads to spin transport effects \cite{Bliokh2005,Torabi2008,Mehrafarin2009,Mehrafarin2010,Mashhadi2010,Mehrafarin2011,
Torabi2012}. We, thus, consider the effect of the spatial fluctuations of the axion field that have survived the inflation, on the spin (polarization) transport of light. Although light wave propagation in inhomogeneous axion field has been  studied in the geometrical optics approximation \cite{Carroll1990,Harari1992}, spin transport effects have been mostly overlooked. Recently,  polarization dependent deflection has been established for light propagating through a scalar cloud surrounding a Kerr black hole \cite{Plascencia}.

Spin transport of electromagnetic and transverse acoustical waves in weakly inhomogeneous media exhibits two principle features, which have been derived using the geometric optics approximation and its adaptations \cite{Bliokh2005,Torabi2008,Mehrafarin2009,Mehrafarin2010,Mashhadi2010,Mehrafarin2011,
Torabi2012}. These features are (i) the rotation of the polarization plane (the Rytov rotation \cite{Rytov1938,Vladimirskii1941}) for linearly polarized waves, which is an example of the geometric Berry phase \cite{Berry1984}, and (ii) the spin Hall (or Magnus) effect for circularly polarized rays, according to which, rays of different polarization propagate along oppositely deflected directions \cite{Dooghin1992,Liberman1992}. Here, we obtain the same effects  for light wave propagation in weakly inhomogeneous axion field via an approach that is apt for the study of spin transport.  By  applying the geometrical optics approximation to the axion electrodynamics field equation, we drive the equation of evolution of circularly polarized states in a flat Friedmann-Robertson-Walker (FRW) background which is relevant to cosmological applications.  This equation, which has a Dirac-like form, is easily solved to yield the phase of the circular waves in the geometrical optics approximation. From this phase factor, we deduce the well known cosmological birefringence effect and drive the differential ray equation for circularly polarized waves, which yields the spin Hall effect. Because the axion field is only weakly inhomogeneous, the deflection of  polarized rays  would be very small.

Although the birefrigence effect has been already established in both uniform and inhomogeneous axion fields \cite{Carroll1990,Carroll1991,Harari1992}, the spin Hall phenomenon is peculiar to inhomogeneous axion field, signalizing small  perturbations in the propagation medium. It may, thus, be used  in the  landscape of axion cosmology, facilitating its progress via the observation of circularly polarized radiation from  CMB as well as other cosmological sources. In this regard, among cosmologically important sources of circular polarization are galaxies, which, like our own Milky Way,  produce polarized synchrotron
radiation by the relativistic circular motion of electrons in the galactic magnetic
field \cite{Enblin2017}. Also, although CMB radiation is expected to be linearly polarized because of anisotropic Compton scattering around the epoch of recombination \cite{Gawiser2000}, there are various physical mechanisms that can induce circular polarization. These include 
primordial magnetic fields \cite{Giovannini2009}, polarized Compton scattering \cite{Vahedi2019}, primordial gravitational waves \cite{Alexander2019}, Faraday conversion by external magnetized plasma \cite{De2015,Cooray2003}, the magnetic field of the so called first stars \cite{King2016}, and, finally, different modifications and symmetry breaking mechanisms beyond the standard model \cite{Zarei2010, Mohammadi2014}. In view of these, recently there has been considerable effort directed towards the determination of upper limits on the observation of circularly polarized CMB radiation \cite{Mainini2013,Nagy2017}.

The primordial origin of axion field inhomogeneity bears an important consequence for the spin transport effects. If the axions appear in the inflationary phase, the quantum field theory that describes them must be invariant under the de Sitter isometry group SO(4,1). Hence, the correlators of the pseudoscalar field, generated by the QFT, are also invariant under the de Sitter group. Now, as is well known from the dS/CFT correspondence (see, e.g., \cite{Antoniadis}), the de Sitter symmetry is approximately realized as $\mathbb{R}^3$ conformal symmetry on the late future boundary, i.e., on the flat spatial cross section of de Sitter space at late future times  (which we take to correspond to the end of inflation). On this boundary, correlations of the field,  constrained by three dimensional conformal invariance, thus depend  only on {\it spatial} separations, which have been stretched to super-Hubble scales due to the exponential expansion. At the end of inflation, we are therefore left with
 small spatial quantum fluctuations of the axion field frozen on super-Hubble scales, the exponential expansion having ironed out large fluctuations. These conformally invariant primordial fluctuations, after re-entering the horizon in the post-inflation era,  constitute the weak long-wavelength inhomogeneity as classical statistical fluctuations of the axion field. (The post-processing of the inhomogeneities due to structure formation is assumed negligible because axions have extremely weak interaction with matter.)  The consequence of this random inhomogeneity for the spin transport  is the fluctuation of the polarization rotation angle and the circularly polarized ray trajectory about their mean uniform field configuration.  Furthermore, the conformal invariance of the  (two-point) correlation function determines the root-mean-square (r.m.s) fluctuations of the trajectory for sufficiently large wave numbers. We point out how the r.m.s fluctuations  can be experimentally determined using data collected from cosmic sources of circularly polarized radiation. Its verification would not only imply the existence of axions but also corroborate that the universe went through a de Sitter stage in accordance with inflation theory.

 The organization of the paper is as follows. In section \ref{sec:geom}, we apply the geometrical optics approximation to axion electrodynamics and drive the equation of evolution of circularly polarized states in a Dirac-like form in the flat FRW universe. In section \ref{sec:spin}, we solve this equation to obtain the phase of the circular waves in the geometrical optics approximation. From this phase factor, we deduce the polarization rotation effect and derive the differential ray equation for circularly polarized waves, which yields the spin Hall effect. These results hold for any weakly inhomogeneous field. In section \ref{sec:primordial}, we consider the implications of the conformally invariant primordial  inhomogeneity on the spin transport effects and finally  conclude by a brief summary in section \ref{sec:summary}.

\section{Evoloution of circularly polarized states in weakly inhomogeneous axion field}\label{sec:geom}

The source-free axion electrodynamics field equation in curved background is ($\hbar=c=1$)
\begin{equation}\label{axEMeqn}
  {\cal D}_\mu (F^{\mu\nu}+g_\phi\phi \tilde{F}^{\mu\nu})=0
\end{equation}
where ${\cal D}_\mu$ represents covariant derivative, $\phi$ is the pseudo-scalar axion field,     $g_\phi$ is the axion-photon coupling with dimension of inverse mass, $F^{\mu\nu}={\cal D}^\mu A^\nu -{\cal D}^\nu A^\mu$  is the Maxwell tensor, and $\tilde{F}^{\mu\nu}=\frac{1}{2}\epsilon^{\mu\nu\rho\sigma}F_{\rho\sigma}$ is its dual, which satisfies ${\cal D}_\mu \tilde{F}^{\mu\nu}=0$. Also, $\epsilon^{\mu\nu\rho\sigma}$ is the complete antisymmetric tensor, which is related to the absolutely antisymmetric Levi-Civita symbol $\varepsilon^{\mu\nu\rho\sigma}$ through the metric according to $\epsilon^{\mu\nu\rho\sigma}=-(\sqrt{-g})^{-1}\varepsilon^{\mu\nu\rho\sigma}$.
Equation \eqref{axEMeqn} is derived by variation of the axion electrodynamics Lagrangian  
\begin{equation}
 \mathcal{L}=-\frac{1}{4}\sqrt{-g}(F_{\mu\nu}F^{\mu\nu}+g_\phi\phi F_{\mu\nu}\tilde{F}^{\mu\nu})
\end{equation}
with respect to the electromagnetic potential  $A_\mu$. In the Lorentz gauge ${\cal D}_\mu A^\mu=0$, \eqref{axEMeqn} becomes
\begin{equation}\label{axEMeqn1}
  \Box A^\nu-\mathcal{R}^\nu_{\ \mu}A^\mu=g_\phi\epsilon^{\nu\mu\sigma\rho}\partial_\mu\phi {\cal D}_\sigma A_\rho,
\end{equation}
where $\Box={\cal D}_\mu{\cal D}^\mu$ and $\mathcal{R}_{\mu\nu}$ is the Ricci tensor.

The long-wavelength inhomogeneity of the axion field implies that it varies insignificantly over the wavelength, $\frac{2\pi}{k}$, of the electromagnetic wave ($g_\phi|\nabla\phi|\ll k$). Hence, we can reliably use the geometrical optics approximation for sufficiently large $k$ and represent the latter
by an almost plane wave ($A^\nu=|A^\nu|\,e^{ik_\mu x^\mu}$) of slowly varying amplitude $|A^\nu|$, viz. ${\cal D}_\mu |A^\nu|\ll k_\mu |A^\nu|$, where $k^\nu=k(1,\hat{\bm{k}})$ is the null vector of the wave ($\hat{\bm{k}}=\bm{k}/k$).
Then, since ${\cal D}_\mu A^\nu \sim i k_\mu A^\nu$, $\Box A^\nu\sim 2i k^\mu{\cal D}_\mu A^\nu$ and $k$ is large,  we can neglect the term $\mathcal{R}^\nu_{\ \mu}A^\mu$ in  \eqref{axEMeqn1} so that the field equation reduces to 
\begin{equation}\label{GOapp}
  k^\mu{\cal D}_\mu A^\nu=\frac{g_\phi}{2}\epsilon^{\nu\mu\sigma\rho}\partial_\mu\phi  k_\sigma A_\rho.
\end{equation}
Also the Lorentz gauge becomes $k_\mu A^\mu=0$. The above is the axion electrodynamics field equation in the geometrical optics approximation.

In the sequel, we consider the spatially flat FRW spacetime
\begin{equation}
  ds^2=R^2(\eta)(-d\eta^2+d\bm{x}^2)
\end{equation}
with scale factor $R$, conformal time $\eta$, and (non-zero) connection coefficients  (no summation over $i=1,2,3$) 
\begin{equation}
\Gamma_{00}^{\, 0}=\Gamma_{ii}^{\, 0}=\Gamma_{i0}^{\, i}=\frac{1}{R}\frac{dR}{d\eta}.
\end{equation}
Calculating the covariant derivatives using these connections and defining $\bm{\mathcal{A}}= R\bm{A}$, we can write \eqref{GOapp} as
\begin{equation}\label{GOvector}
    (\partial_\eta+\hat{\bm{k}}\cdot\nabla)\bm{\mathcal{A}}=\frac{g_\phi}{2} (\partial_\eta\phi\hat{\bm{k}}+\nabla\phi)\times\bm{\mathcal{A}}-(\frac{g_\phi}{2}\nabla\phi\times\hat{\bm{k}}+\frac{R^\prime}{R}\hat{\bm{k}})\,\hat{\bm{k}}\cdot\bm{\mathcal{A}} 
  \end{equation}
 Note that the above is the spatial component of equation \eqref{GOapp}, its zero component is not independent because of the Lorentz gauge condition  (${\mathcal A}^0=\hat{\bm{k}}\cdot\bm{\mathcal A}$) and corresponds to just taking the scalar product of \eqref{GOvector} with $\hat{\bm{k}}$. In terms of the parallel and normal (to $\hat{\bm{k}}$) components of $\bm{\mathcal A}$ viz. $ \bm{\mathcal A}_\|=\hat{\bm{k}}\,(\hat{\bm{k}}\cdot\bm{\mathcal{A}})$ and $ \bm{\mathcal A}_\perp=\bm{\mathcal A}-\bm{\mathcal A}_\|$, the right hand side of \eqref{GOvector} reads:
$$
\frac{g_\phi}{2} (\partial_\eta\phi\hat{\bm{k}}+\nabla\phi)\times\bm{\mathcal{A}}_\perp-\frac{R^\prime}{R}\bm{\mathcal{A}}_\|
$$
so that the equation can be decomposed as
\begin{eqnarray}\begin{array}{c}
    (\partial_\eta+\hat{\bm{k}}\cdot\nabla)\bm{\mathcal{A}}_\perp=\frac{g_\phi}{2} [\partial_\eta\phi\,\hat{\bm{k}}\times\bm{\mathcal{A}}_\perp+(\nabla\phi\times\bm{\mathcal{A}}_\perp)_\perp] \\
(\partial_\eta+\hat{\bm{k}}\cdot\nabla)(R\bm{\mathcal{A}}_\|)=\frac{g_\phi}{2}(\nabla\phi\times R\bm{\mathcal{A}}_\perp)_\|.\label{decom}
\end{array}
\end{eqnarray}
The first equation determines $\bm{\mathcal{A}}_\perp$, which then yields $\bm{\mathcal{A}}_\|$ through the second equation.  Thus there are two independent components given by $\bm{\mathcal{A}}_\perp$, but notice from the second equation that the freedom in setting $\bm{\mathcal{A}}_\|$ equal to zero (which corresponds to the Coulomb gauge $A^0=0$) is lost because of the inhomogeneity of the axion field.

$\bm{\mathcal{A}}_\perp$ define the circularly polarized states $\psi_\pm$ (taking the $z$ axis along $\bm{k}$,  $\psi_\pm=\mp\mathcal{A}_x+i\mathcal{A}_y$). 
Expressed as equation for polarized states, the first of equations \eqref{decom} takes the Dirac-like form
\begin{equation}\label{finalpol}
  i\partial_\eta\psi_\sigma=H\psi_\sigma,
\end{equation}
where $\sigma=\pm$ represents right/left polarization and the Hamiltonian operator is given by
\begin{equation}\label{finalhamiltoni}
 {H}=\hat{\bm{k}}\cdot{\bm{P}}+\frac{g_\phi}{2}\sigma(\partial_\eta\phi +\hat{\bm{k}}\cdot\nabla\phi)
\end{equation}
$\bm{P}=-i\nabla$ being the momentum operator.

Equation \eqref{finalpol} is the equation of evolution of polarized states of axion electrodynamics in the geometrical optics approximation and holds for any weakly inhomogeneous axion field in a flat FRW background.

\section{Polarization rotation and the spin Hall effect}\label{sec:spin}

Using equation \eqref{finalpol}, we can establish the following spin transport effects in any weakly inhomogeneous axion field.  Solving \eqref{finalpol} yields 
$ \psi_\sigma \propto e^{iS}$
where 
\begin{equation}\label{phase}
    S(\eta,\bm{x})= -k\eta+\bm{k}\cdot\bm{x}-\frac{g_\phi}{2}\sigma\phi(\eta,\bm{x})
\end{equation}
is the phase of the circular wave. The first two terms constitute the plane wave front, while the last term gives the slowly spatially varying sinusoidal amplitude of the wave $\bm{A}$ as
$
\frac{1}{R}(\hat{\bm{\imath}} \sin \frac{g_\phi\phi}{2}-\hat{\bm{\jmath}} \cos\frac{g_\phi\phi}{2}),
$
where $\hat{\bm{\imath}},\hat{\bm{\jmath}},\hat{\bm{k}}$  form an orthonormal set. This corresponds to the local rotation of the polarization plane by $g_\phi\phi/2$, a result previously, albeit differently, obtained for inhomogeneous axion field (cosmological birefringence) \cite{Carroll1990,Harari1992}.

The inhomogeneity of the axion field also affects the ray equation, as is seen from \eqref{phase}. The rays are perpendicular to the wavefronts $S=\text{const.}$ and   their trajectory is given by the ray differential equation 
\begin{equation}
  \frac{d\bm{x}}{ds}=\frac{\nabla S}{|\nabla S|}
\end{equation}
where $ds=|d\bm{x}|$ is the differential path length. In uniform axion field, $ d\bm{x}/ds=\hat{\bm{k}}$, corresponding to plane wave. In presence of inhomogeneity, however,  we have from \eqref{phase} to first order in $g_\phi|\nabla\phi|/k\ll 1$,
\begin{equation}
  \frac{d\bm{x}}{ds}=\hat{\bm{k}}-\frac{g_\phi\sigma}{2k}\nabla_\perp \phi 
\end{equation}
 where $\nabla_\perp =\nabla-\hat{\bm{k}}(\hat{\bm{k}}\cdot\nabla)$ is the component of $\nabla$ perpendicular to $\hat{\bm{k}}$. The first term is the standard zero-order term in geometrical optics approximation, while the second term yields the spin Hall effect. Accordingly, the left and right circularly polarized rays oppositely deflect from the zero-order trajectory, the transverse  (local) deflections, $\frac{g_\phi\sigma}{2k} \nabla_\perp \phi  ds$, having a small  magnitude ($g_\phi|\nabla\phi|\ll k$).

\section{Conformally invariant primordial inhomogeneity}\label{sec:primordial}

The inhomogeneity, which is inherited from the primordial quantum fluctuations, corresponds to small perturbations in an otherwise uniform field, $\varphi(\eta)$. We, therefore, write
\begin{equation}\label{perturb}
\phi(\eta, \bm{x})=\varphi(\eta)+\phi_0(\bm{x})
\end{equation}
where, the long-wavelength spatial perturbation $\phi_0$ is traced back to the super Hubble fluctuations of the axion field at the end of inflation that have subsequently re-entered the horizon. It has zero mean and its correlator  is fixed by conformal invariance to be 
\begin{equation}\label{correl}
\left\langle \phi_0(\bm{x})\phi_0(\bm{x}^\prime)\right\rangle =H^2\left(\frac{H^{-1}}{|\bm{x}-\bm{x}^\prime|}\right)^{2\Delta}
\end{equation}
where $|\bm{x}-\bm{x}^\prime|> H^{-1}$ and $\Delta=\frac{3}{2}\pm\sqrt{\frac{9}{4}-\frac{m_\phi^2}{H^2}}\approx \frac{m_\phi^2}{3H^2}$ is the scaling dimension of the axion field, $m_\phi$ being its mass, and $H$, the Hubble constant of the inflationary phase. Using the order of magnitude estimates $m_\phi\sim10^{-5}\text{eV}$ (suitable for dark matter \cite{Klaer}) and $H\sim 10^{34}\text{s}^{-1}$, gives $\Delta\sim 10^{-48}$. Equation \eqref{correl} follows from the standard result for the equal time correlator in de Sitter spacetime \cite{Antoniadis}
\begin{equation}
\left\langle \phi(\eta, \bm{x})\phi(\eta, \bm{x}^\prime)\right\rangle=H^2 \left(\frac{\eta}{|\bm{x}-\bm{x}^\prime|}\right)^{2\Delta}
\end{equation}
where $\eta \ll  |\bm{x}-\bm{x}^\prime|$, upon writing $\eta\sim H^{-1}$ corresponding to the end of inflation. Then, $|\bm{x}-\bm{x}^\prime|> H^{-1}$, reconciling with super Hubble scale correlations.

\begin{figure}[h]
  \centering
  \begin{tikzpicture}[x={(0cm,1cm)}, y={(-0.866cm,-0.5cm)}, z={(0.866cm,-0.5cm)}, scale=1.5,
    >=stealth, 
    inner sep=0pt, outer sep=2pt,
    plate/.style={fill=black!50!white, opacity=0.3}, ]
    \colorlet{darkgreen}{green!50!black}
    \colorlet{lightgreen}{green!50!white}
    \colorlet{darkred}{red!50!black}
    \colorlet{lightred}{red!40!white}
        \coordinate (O) at (0, 0, 0);
    \draw[dashed] (O) -- (0, 0,  4);
    \draw[ultra thick,violet,->] (0,0,-0.7) -- (O);
    \draw (0,0,-0.2) node [violet,above right] {$\bm{k}$};
      \filldraw[plate] (-0.15,-0.5,3) -- (-0.5,0.5,3) -- (0.5,0.5,3) -- (0.85,-0.5,3) -- (-0.15,-0.5,3); 
\filldraw[plate] (-0.15,-0.3,2) -- (-0.5,0.7,2) -- (0.5,0.7,2) -- (0.85,-0.3,2) -- (-0.15,-0.3,2);
      \pgfmathsetmacro{\xcoora}{0.18}
      \pgfmathsetmacro{\ycoora}{-0.15}
      \coordinate (coordinatea positive) at (\xcoora,\ycoora,1);
      \coordinate (coordinatea negative) at (-\xcoora,-\ycoora,1);
      \pgfmathsetmacro{\xcoorb}{-0.35}
      \pgfmathsetmacro{\ycoorb}{0.05}
      \coordinate (coordinateb positive) at (\xcoorb,\ycoorb,1.8);
      \coordinate (coordinateb negative) at (-\xcoorb,-\ycoorb,1.8);
      \pgfmathsetmacro{\xcoorc}{0.1}
      \pgfmathsetmacro{\ycoorc}{-0.25}
      \coordinate (coordinatec positive) at (\xcoorc,\ycoorc,3);
      \coordinate (coordinatec negative) at (-\xcoorc,-\ycoorc,3);
      \pgfmathsetmacro{\xcoord}{-0.07}
      \pgfmathsetmacro{\ycoord}{-0.14}
      \coordinate (coordinated positive) at (\xcoord,\ycoord,3.5);
      \coordinate (coordinated negative) at (-\xcoord,-\ycoord,3.7);
          \draw[thick,blue,->] plot [smooth] coordinates {(0,0,0) (coordinatea positive) (coordinateb positive)(coordinatec positive)(coordinated positive)};
      \draw[thick,darkred,->] plot [smooth] coordinates {(0,0,0) (coordinatea negative) (coordinateb negative)(coordinatec negative)(coordinated negative)};
      \draw [blue,thick,->] (\xcoora-0.05,\ycoora+0.05,1) arc (275:-10:4pt);
      \draw (\xcoora +0.15,\ycoora +0.15,0.7) node [blue, above right] {$\sigma=-$};
      \draw [darkred,thick,->] (-\xcoora+0.05,-\ycoora-0.05,1) arc (-10:275:4pt);
      \draw (-\xcoora0,-\ycoora0,1.3) node [darkred, below left] {$\sigma=+$};
\draw[darkgreen,->] (coordinateb negative) -- (coordinateb positive); 
      \draw[darkgreen,->] (coordinatec negative) -- (coordinatec positive);
      \draw (coordinatec positive) node [darkgreen, above right] {$g_\phi\nabla_\perp\phi_0/k$};
            \end{tikzpicture}
\caption{A schematic picture of spin Hall effect. $\nabla_\perp\phi_0$  fluctuates in the (depicted) transverse plane so that left and right polarized rays oppositely fluctuate about the mean uniform field propagation axis $\bm{k}$, while evolving in that direction.} \label{fig1}
\end{figure}
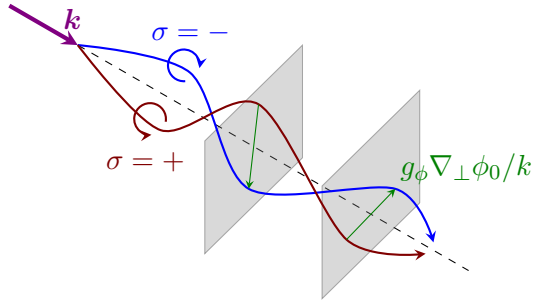

In view of \eqref{perturb}, the polarization rotation angle fluctuates in space about the uniform field value $g_\phi\varphi/2$. Also, the first order ray equation 
\begin{equation}\label{ray}
  \frac{d\bm{x}}{ds}=\hat{\bm{k}}-\frac{g_\phi\sigma}{2k}\nabla_\perp \phi_0
\end{equation}
 is a stochastic equation with correlated noise.
The left and right circularly polarized rays, therefore, oppositely fluctuate about the mean (uniform field) propagation axis $\bm{k}$ while evolving in that direction  (FIG. \ref{fig1}).  

Taking the $z$-axis along $\bm{k}$, from \eqref{ray} we have $s=z$ to first order, and 
\begin{equation}
\bm{x}_\perp (z)=-\frac{g_\phi\sigma}{2k}\int_{0}^z  dz^\prime [\nabla_\perp^\prime\phi_0(\bm{x}^\prime)]_{\bm{x}_\perp^\prime=0}
\end{equation}
where the ray is assumed to have originated at $z=0$ and the integrand is evaluated at $\bm{x}_\perp^\prime=0$ because $s=z$ on the trajectory. In the absence of a definite model for $\phi_0$, one cannot obtain the deflections of circularly polarized rays from the above equation. However, using the correlator \eqref{correl} which is fixed by symmetry and model independent, the mean square deflections can be calculated furnishing our basic observable. Thus, from \eqref{correl}, for the mean square fluctuations at distant $z$ from the source  we have the first order result
\begin{eqnarray}
\left\langle  \bm{x}_\perp^2(z) \right\rangle&=&-\frac{k^{-2}}{4}\frac{g_\phi^2}{H^{2(\Delta-1)}}\int_{0}^z\! \int_{0}^z dz^\prime dz^{\prime\prime}\left[\nabla_\perp^{\prime 2}|\bm{x}^\prime-\bm{x}^{\prime\prime}|^{-2\Delta}\right]_{\bm{x}_\perp^\prime=\bm{x}_\perp^{\prime\prime}=0} \nonumber \\&=&k^{-2}\frac{\Delta g_\phi^2}{H^{2(\Delta-1)}}\int_0^z\! \int_0^z \frac{dz^\prime dz^{\prime\prime}}{(z^\prime-z^{\prime\prime})^{2(\Delta+1)}}
\end{eqnarray}
where $|z^\prime- z^{\prime\prime}|> H^{-1}$. Changing the integration variable to $u=z^\prime+z^{\prime\prime},\, v=z^\prime-z^{\prime\prime}$, we obtain 
\begin{equation}
\left\langle  \bm{x}_\perp^2(z) \right\rangle=k^{-2}\frac{\Delta g_\phi^2}{H^{2(\Delta-1)}}\int _{v=\frac{1}{H}}^{\frac{z}{\surd{2}}}\!\int_{u=v}^{\surd{2}z-v} \! v^{-2(\Delta+1)}du\,dv=\left (\frac{2\Delta+\zeta^{-(2\Delta+1)}}{2\Delta+1}\zeta-1\right)g_\phi^2H^2k^{-2}
\end{equation}
where $\zeta=zH/\surd{2}>1$. Therefore, expanding to first order in $\Delta\approx 0$,
\begin{equation}\label{var}
\left\langle  \bm{x}_\perp^2(z) \right\rangle=2\Delta (\zeta-\ln \zeta-1)g_\phi^2H^2k^{-2}\approx \frac{\surd{2}}{3}m_\phi^2g_\phi^2Hzk^{-2}
\end{equation}
where the last expression holds for large $\zeta$. Using the experimental bound $g_\phi<8.8\times 10^{-20}\, \text{eV}^{-1}$ for $m_\phi<0.02\, \text{eV}$ \cite{Andria} and the above mentioned estimates, for $z\approx 3/\surd{2}g_\phi^2m_\phi^2H$, which is of the order of megaparsecs, we have $\left\langle  \bm{x}_\perp^2 \right\rangle\approx  k^{-2}$
so that  the r.m.s fluctuations is comparable to the wavelength (for sufficiently small wavelengths).

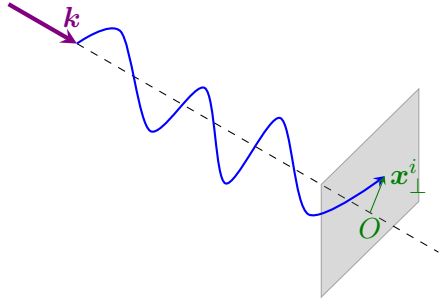
\begin{figure}[h]
  \centering
  \begin{tikzpicture}[x={(0cm,1cm)}, y={(-0.866cm,-0.5cm)}, z={(0.866cm,-0.5cm)}, scale=1.5,
    >=stealth, 
    inner sep=0pt, outer sep=2pt,
    plate/.style={fill=black!50!white, opacity=0.3}, ]
    \colorlet{darkgreen}{green!50!black}
    \colorlet{lightgreen}{green!50!white}
    \colorlet{darkred}{red!50!black}
    \colorlet{lightred}{red!40!white}
        \coordinate (O) at (0, 0, 0);
    \draw[dashed] (O) -- (0, 0,  3.7);
    \draw[ultra thick,violet,->] (0,0,-0.7) -- (O);
    \draw (0,0,-0.2) node [violet,above right] {$\bm{k}$};
      \filldraw[plate] (-0.15,-0.5,3) -- (-0.5,0.5,3) -- (0.5,0.5,3) -- (0.85,-0.5,3) -- (-0.15,-0.5,3); 
      \pgfmathsetmacro{\xcoora}{0.18}
      \pgfmathsetmacro{\ycoora}{-0.15}
      \coordinate (coordinatea positive) at (\xcoora,\ycoora,.3);
          \pgfmathsetmacro{\xcoorb}{-0.35}
      \pgfmathsetmacro{\ycoorb}{0.05}
      \coordinate (coordinateb positive) at (\xcoorb,\ycoorb,.8);
  \pgfmathsetmacro{\xcoorg}{0.15}
      \pgfmathsetmacro{\ycoorg}{-0.11}
      \coordinate (coordinateg positive) at (\xcoorg,\ycoorg,1.2);
          \pgfmathsetmacro{\xcoorh}{-0.4}
      \pgfmathsetmacro{\ycoorh}{0.08}
      \coordinate (coordinateh positive) at (\xcoorh,\ycoorh,1.6);
\pgfmathsetmacro{\xcoori}{0.2}
      \pgfmathsetmacro{\ycoori}{-0.18}
      \coordinate (coordinatei positive) at (\xcoori,\ycoori,1.9);
          \pgfmathsetmacro{\xcoorj}{-0.3}
      \pgfmathsetmacro{\ycoorj}{0.03}
      \coordinate (coordinatej positive) at (\xcoorj,\ycoorj,2.4);
      \pgfmathsetmacro{\xcoorc}{0.15}
      \pgfmathsetmacro{\ycoorc}{-0.25}
      \coordinate (coordinatec positive) at (\xcoorc,\ycoorc,2.9);
 \coordinate (coordinatec negative) at (0,0,3);
              \draw[thick,blue,->] plot [smooth] coordinates {(0,0,0) (coordinatea positive) (coordinateb positive)(coordinateg positive)(coordinateh positive)(coordinatei positive)(coordinatej positive)(coordinatec positive)};
            \draw[darkgreen,->] (coordinatec negative) -- (coordinatec positive); 
\draw (coordinatec positive) node [darkgreen, right] {$\bm{x}_\perp^i$};
\draw (coordinatec negative) node [darkgreen, below] {$O$};
 \end{tikzpicture}
     \caption{Experimental determination of ray path fluctuations. $\bm{x}_\perp^i$ represents the  point where the fluctuating circularly polarized ray from a cosmic source hits the (depicted) transverse photographic plate at time $i$. Over time, a large data set can be gathered of such points, enabling the standard deviation to be determined.} \label{fig2}
\end{figure}

Experimentally, the r.m.s fluctuations given by \eqref{var} corresponds to the standard deviation of a large data sample $\{\bm{x}_\perp^i\}$ collected over time from a cosmically distant source of circularly polarized radiation, $\bm{x}_\perp^i$ being the position of the point (with respect to the origin, $\sum_i \bm{x}_\perp^i=0$) where the fluctuating ray (of a given wavelength) hits the transverse `photographic plate' at time $i$ (FIG. \ref{fig2}). Whence, by the log-log plot of the standard deviation (obtained from samples with different wavelengths) versus wavelength,  the prediction can be tested. 

\section{Summary}\label{sec:summary}
We have considered light wave propagation through weakly inhomogeneous axion field in flat FRW background using geometrical optics approximation by an approach that is apt for the study of spin transport. We have established the spin transport effects consisting in the rotation of polarization plane (cosmological birefringence) for linearly polarized waves and the spin Hall effect for circularly polarized rays. The latter prediction, which is peculiar to inhomogeneous axion field and signalizes small perturbations in the field, may be experimentally employed via the observation of circularly polarized cosmic radiation, facilitating progress in observational cosmology.

The inhomogeneity could have been inherited from the primordial quantum fluctuations of the axion field that have survived the inflation. At the end of inflation, these fluctuations become super horizon and conformally invariant, and after again becoming subhorizon in the post-inflation era, they constitute the weak long-wavelength inhomogeneity  as classical statistical fluctuations of the axion field. As a consequence of this random inhomogeneity, the polarization rotation angle fluctuates about its mean uniform field value. Also, the left and right polarized ray trajectories oppositely fluctuate about the uniform field propagation axis, and, as we have shown, the conformally invariant correlator determines the r.m.s fluctuations for sufficiently large wave numbers. We have explained how the r.m.s fluctuations can be experimentally determined using data collected from cosmic sources of circularly polarized radiation.  Verification of this prediction would corroborate the existence of axions and inflation theory at the same time.

\end{document}